\documentclass[prb,preprint]{revtex4-1} 

\usepackage{amsmath}  
\usepackage{amsfonts} 
\usepackage{graphicx} 

\graphicspath{%
    {converted_graphics/}
    {/}
}
\begin{document}


\title{An Effective Photon Momentum in a Dielectric Medium:\\ A Relativistic Approach}

\author{Bradley W.\ Carroll}
\author{Farhang Amiri}
\author{J.\ Ronald Galli}

\affiliation{Department of Physics, Weber State University, Ogden, UT 84408}


\date{\today}

\begin{abstract}
We use a relativistic argument to define an effective photon that travels through a transparent (non-absorbing) nondispersive dielectric medium of index of refraction $n$. If $p$ is the momentum of the photon in a vacuum, then the momentum of an effective photon inside the medium may be of the form $p_{\rm eff}=pn^\alpha$ and still reproduce the observed transverse relativistic drift when the medium is in motion.  We employ an energy argument to determine the value of the exponent to be $\alpha =-1$, so that the effective photon momentum is $p_{\rm eff}=p/n$, which is the Abraham momentum.
\end{abstract}

\maketitle 

\section{Introduction} 

The refraction of light as it passes from a vacuum into a transparent dielectric medium and undergoes a speed change is well-established and is included in standard electromagnetism texts.\cite{Griffiths}  On a microscopic level, it is understood that the propagation of a light ray through a medium is actually a complicated process that involves multiple scatterings of the incident electromagnetic wave by the atoms that comprise the bulk of the medium.\cite{Feynman}  The refracted and reflected rays are the result of the interference of these scattered waves.  The net result of this interference is that the refracted ray moves with a reduced speed of $c/n$ inside a medium with index of refraction $n$.  The wave's Poynting vector $\vec{S}$ is directed along the path followed by the ray.

The situation is more complicated when the refracted ray is described from the point of view of its constituent photons.  First, we consider the case of a single photon that is incident upon a transparent dielectric block at rest.  The photon has a certain probability of being transmitted or reflected from the block's surface.  We will restrict our analysis to the case when the photon is transmitted into the block.  Photons are massless particles and, as such, must travel at the speed of light; hence the refracted ray, moving at $c/n$, cannot be composed of massless photons.  Instead, we will define an {\em effective photon} that follows the path of the refracted ray, transports the ray's momentum and energy, and moves with speed $c/n$.  The mass of the effective photon is denoted by $m$, which we take to be Lorentz-invariant.  We define the effective photon's energy by the usual relativistic expression
\begin{equation}
E_{\rm eff}=\gamma_v mc^2
\label{eq:relE}
\end{equation}
and the magnitude of its momentum by
\begin{equation}
p_{\rm eff}=\gamma_v mv,
\label{eq:relp}
\end{equation}
where $\gamma_v$ is the Lorentz factor for the effective photon's velocity,
\begin{equation}
\gamma_v=\dfrac{1}{\sqrt{1-\dfrac{v^2}{c^2}}}=\dfrac{1}{\sqrt{1-\dfrac{1}{n^2}}}.
\label{eq:gammav}
\end{equation}
As usual, these may be combined to yield
\begin{equation}
E_{\rm eff}^2-p_{\rm eff}^2c^2=m^2c^4,
\label{eq:Epm}
\end{equation}
which expresses the invariant magnitude of the effective photon's momentum-energy four vector.  Substituting $v=c/n$ for the effective photon's speed, Eq.~(\ref{eq:relp}) becomes
\begin{equation}
p_{\rm eff}=\gamma_v\frac{mc}{n},
\label{eq:relp2}
\end{equation}
and Eq.~(\ref{eq:relE}) shows that the effective photon's momentum and energy are related by 
\begin{equation}
E_{\rm eff}=np_{\rm eff}c
\label{eq:Enpc}
\end{equation}
Substituting this into Eq.~(\ref{eq:Epm}) shows that the mass of the effective photon is
\begin{equation}
m=\frac{p_{\rm eff}}{c}\sqrt{n^2-1}.
\label{eq:mass}
\end{equation}

We assume that the momentum of the effective photon inside a dielectric medium at rest is related to the momentum $p$ of the incident momentum in vacuum by 
\begin{equation}
p_{\rm eff}=n^\alpha p,
\label{eq:effp}
\end{equation}
where $\alpha$ is a real number.  The energy of the effective photon is then, from Eq.~(\ref{eq:Enpc}),
\begin{equation}
E_{\rm eff}=n^{\alpha +1}pc.
\label{eq:effE}
\end{equation}
The value of $\alpha$ has not yet been determined.  The two possibilities most often discussed are $\alpha =1$ so $p_{\rm eff}=np$ (the so-called Minkowski momentum) and $\alpha =-1$ so $p_{\rm eff}=p/n$ (the so-called Abraham momentum).\cite{Barnett2010} 

\section{The Transverse Relativistic Drift}
    
To investigate the implications of defining an effective photon in this way, we will utilize the phenomenon of the ``aether drift,'' the change in the velocity of light in a moving medium predicted by A.\ Fresnel\cite{Fresnel} in 1818 and first detected experimentally by H.\ Fizeau\cite{Fizeau} in 1851.  This effect is now understood to be entirely due to the relativistic transformation of the velocity of light from its value in the rest frame of the medium.  For this reason, we will refer to this change in velocity as a ``relativistic drift.''

Consider a rectangular block of transparent nondispersive dielectric material of thickness $T$ that is at rest relative to an inertial reference frame $S'$, which we call the ``rest frame.''  We restrict ourselves to the special case where the top and bottom faces of the block are parallel to the direction of motion and perpendicular to the $y'$ axis.  The frame $S'$ moves in the positive $x$-direction with velocity $u$ relative to another inertial reference frame $S$, which we call the ``lab frame.''  Thus an observer at rest in the lab frame will see the dielectric block moving in the positive $x$-direction with velocity $u$ (see Fig.~\ref{fig:refFrames}).  We stipulate that the block remains at rest in frame $S'$ when the photon enters because the block is sufficiently massive or is otherwise constrained.  


Gjurchinovski\cite{Gjurchinovski} used the relativistic velocity transformations to show that the angles of incidence ($\theta_1$) and refraction ($\theta_2$) of the ray in the moving block are related in the lab frame $S$ by
\begin{equation}
\tan\theta_2=\gamma\,\dfrac{\left(n^2-1\right)\dfrac{u}{c}+\left(1-\dfrac{n^2u^2}{c^2}\right)\sin\theta_1}{\sqrt{n^2\left(1-\dfrac{u}{c}\sin\theta_1\right)^2-\left(\sin\theta_1-\dfrac{u}{c}\right)^2}},
\label{eq:tanth2v}
\end{equation}
where $\gamma=1/\sqrt{1-u^2/c^2}$.  (The ``1'' subscript denotes a value in the vacuum below the dielectric block, and the ``2'' subscript denotes a value within the dielectric block.)  If we set $u=0$, it becomes equivalent to the familiar Snell's law, $\sin\theta_1=n\sin\theta_2$.

The significance of Eq.~(\ref{eq:tanth2v}) is seen for the case of a normally incident ray in the frame $S$.  Setting $\theta_1=0$ produces
\begin{equation}
\tan\theta_2=\gamma\,\dfrac{\left(n-\dfrac{1}{n}\right)\dfrac{u}{c}}{\sqrt{1-\dfrac{u^2}{n^2c^2}}}.
\label{eq:drift}
\end{equation}
For non-zero $u$, this means that, in the lab frame $S$ for normal incidence ($\theta_1=0$), the refracted ray deviates slightly in the positive $x$-direction.  If $u/c\ll 1$, then to first order in $u/c$ for a block of thickness $T$, the ray experiences a transverse relativistic drift $\Delta x$ of
\begin{equation}
\Delta x=T\tan\theta_2\simeq\left(n-\dfrac{1}{n}\right)\dfrac{Tu}{c}.
\end{equation}
This is Fresnel's formula for his ``aether drift.''  A transverse relativistic drift has been experimentally confirmed by Jones\cite{Jones1971,Jones1972,Jones1975} and Leach {\em et al.}.\cite{Leach}

\section{Lorentz Transformation of the Wave Four-Vector}

\begin{figure}[h] 
  \centering
  \includegraphics[bb=0 0 548 171,width=5.67in,height=1.76in,keepaspectratio]{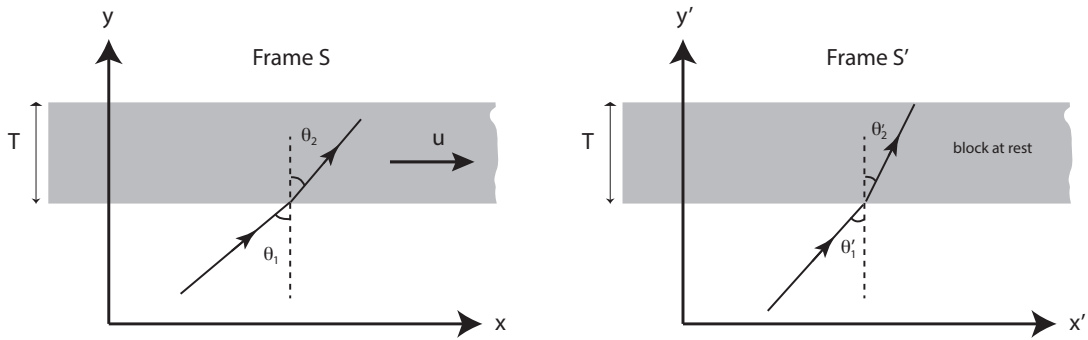}
  \caption{The reference frames used in this paper.  Relative to reference frame $S$ (the lab frame), the dielectric block and reference frame $S'$ both move in the positive $x$-direction with speed $u$.  The block remains at rest in frame $S'$ (the rest frame). 
The situation shown is for $n=1.5$ and $u/c=0.2$.}
  \label{fig:refFrames}
\end{figure}

To understand better the nature of light in a moving dielectric, we study the same phenomenon by performing a Lorentz transformation of the wave four-vector, $(k_x, k_y, k_z, \omega/c)$.  Specifically, as shown in Fig.~\ref{fig:refFrames}, we will consider a ray of light that is incident on the moving transparent nondispersive dielectric block at a certain angle of incidence $\theta_1'$.  We will then transform to a reference frame in which the block is at rest.  In that reference frame, we use Snell's law,
\begin{equation}
\sin\theta_1'=n\sin\theta_2',
\label{eq:Snell}
\end{equation}
and well-established boundary conditions to describe wave four-vector after it enters the block.  Finally, we transform back to the reference frame in which the block is moving, and calculate the refracted wave four-vector in the moving block as a function of the original angle of incidence.  The relevant Lorentz transformations are
\begin{eqnarray}
k_x'&=&\gamma\left(k_x-\dfrac{u\omega}{c^2}\right)\\[1ex]
\omega'&=&\gamma\left(\omega-uk_x\right).
\end{eqnarray} 
We begin with a ray of light with wavenumber $k_1=2\pi/\lambda_1$ and angular frequency $\omega_1$ as observed in the lab frame $S$.  The ray is incident upon the bottom surface of the block, making an angle $\theta_1$ with the normal to the block's surface.  The $x$-component of the wave vector is
\begin{equation}
k_{1x}=k_1\sin\theta_1.\label{eq:k1x}
\end{equation}    
We now use the Lorentz transformations to find the components of the incident wave vector in frame $S'$.  Employing the fact that in the vacuum
\begin{equation}
\dfrac{\omega_1}{k_1}=\dfrac{\omega_1'}{k_1'}=c\label{eq:omoverk},
\end{equation}
we obtain
\begin{eqnarray} 
k_{1x}'&=&\gamma\left(k_{1x}-\dfrac{u\omega_1}{c^2}\right)=\gamma\left(k_1\sin\theta_1-\dfrac{uk_1c}{c^2}\right)=\gamma k_1\left(\sin\theta_1-\dfrac{u}{c}\right)\\[1ex]
\omega_1'&=&\gamma\left(\omega_1-uk_{1x}\right)=\gamma\left(k_1c-uk_1\sin\theta_1\right)=\gamma k_1c\left(1-\dfrac{u}{c}\sin\theta_1\right).\label{eq:omega1p}
\end{eqnarray}

Because the dielectric block is at rest in frame $S'$, the wavelength $\lambda_2'$ inside the block is related to $\lambda_1'$ by $\lambda_2'=\lambda_1'/n$.  Then the wave number $k_2'$ within the block at rest is
\begin{equation}
k_2'=\dfrac{2\pi}{\lambda_2'}=\dfrac{2\pi}{\lambda_1'/n}=nk_1'.
\end{equation}
Combining this with Snell's law, Eq.~(\ref{eq:Snell}), we have
\begin{equation}
k_2'\sin\theta_2'=\left(nk_1'\right)\left(\dfrac{1}{n}\sin\theta_1'\right)=k_1'\sin\theta_1'
\end{equation}
or
\begin{equation}
k_{2x}'=k_{1x}'=\gamma k_1\left(\sin\theta_1-\dfrac{u}{c}\right).
\end{equation}
Because the angular frequency does not change upon entering the block in the rest frame $S'$, we have
\begin{equation}
\omega_2'=\omega_1'=\gamma k_1c\left(1-\dfrac{u}{c}\sin\theta_1\right).
\end{equation}
Thus
\begin{equation}
k_1'=\dfrac{\omega_1'}{c}=\gamma k_1\left(1-\dfrac{u}{c}\sin\theta_1\right).
\end{equation}
The $y'$-component of the wave vector within the block in frame $S'$ is obtained from
\begin{equation}
k_{2y}'=\sqrt{k_2'^2-k_{2x}'^2}=\sqrt{\left(nk_1'\right)^2-\left[\gamma k_1\left(\sin\theta_1-\dfrac{u}{c}\right)\right]^2}.
\end{equation}
Therefore
\begin{equation}
k_{2y}'=\gamma k_1\sqrt{n^2\left(1-\dfrac{u}{c}\sin\theta_1\right)^2-\left(\sin\theta_1-\dfrac{u}{c}\right)^2}.
\end{equation}

Finally, we use the inverse transformations,
\begin{eqnarray}
k_{2x}&=&\gamma\left(k_{2x}'+\dfrac{u\omega_2'}{c^2}\right)\\[1ex]
k_{2y}&=&k_{2y}'
\end{eqnarray} 
to obtain the components of the wave four-vector inside the moving block as measured from the lab frame $S$.  We are really interested in $\tan\theta_2$,
\begin{equation}
\tan\theta_2=\dfrac{k_{2x}}{k_{2y}}=\gamma\dfrac{k_{2x}'+\dfrac{u\omega_2'}{c^2}}{k_{2y}'}.
\end{equation}
Thus
\begin{equation}
\tan\theta_2=\gamma\,\dfrac{\left(\sin\theta_1-\dfrac{u}{c}\right)+\dfrac{u}{c}\left(1-\dfrac{u}{c}\sin\theta_1\right)}{\sqrt{n^2\left(1-\dfrac{u}{c}\sin\theta_1\right)^2-\left(\sin\theta_1-\dfrac{u}{c}\right)^2}}
\end{equation}
so that
\begin{equation}
\tan\theta_2=\dfrac{\sin\theta_1}{\gamma \sqrt{n^2\left(1-\dfrac{u}{c}\sin\theta_1\right)^2-\left(\sin\theta_1-\dfrac{u}{c}\right)^2}}.\label{eq:tanth2k}
\end{equation}

We thus find that the directions of the light ray (Eq.~\ref{eq:tanth2v}) and the wave vector (Eq.~\ref{eq:tanth2k}) in the moving block are not the same.  In particular, for normal incidence ($\theta_1=0$) in the lab frame $S$, Eq.~(\ref{eq:tanth2k}) shows that $\tan\theta_2=0$, so $\theta_2=0$.  That is, while the ray's velocity vector shows a transverse relativistic drift (Eq.~\ref{eq:drift}), the wave vector does not display a transverse relativistic drift.  The ray's velocity vector is in the same direction as its Poynting vector $\vec{S}$, so we must conclude that the directions of $\vec{S}$ and $\vec{k}$ diverge in the moving block. 

This divergence of the directions of $\vec{S}$ and $\vec{k}$ is caused by the directional-dependence of the ray's velocity inside the moving block.  For the block at rest in frame $S'$, the components of the ray's velocity within the dielectric block are
\begin{equation}
v_{2x}'=\dfrac{c}{n}\sin\theta_2'
\label{eq:v2xp}
\end{equation}
and
\begin{equation}
v_{2y}'=\dfrac{c}{n}\cos\theta_2'.
\label{eq:v2yp}
\end{equation}
The velocity transformation equations provide the components of the ray's velocity inside the moving block:
\begin{equation}
v_{2x}=\dfrac{v_{2x}'+u}{1+\dfrac{uv_{2x}'}{c^2}}=\dfrac{\dfrac{c}{n}\sin\theta_2'+u}{1+\dfrac{u}{cn}\sin\theta_2'},
\label{eq:vxtransinv}
\end{equation}
and
\begin{equation}
v_{2y}=\dfrac{v_{2y}'}{\gamma\left(1+\dfrac{uv_{2x}'}{c^2}\right)}=\dfrac{\dfrac{c}{n}\cos\theta_2'}{\gamma\left({1+\dfrac{u}{cn}\sin\theta_2'}\right)}.
\label{eq:vytransinv}
\end{equation}    
Obviously, the magnitude of the velocity of the ray in the moving block is not constant, but varies with direction.  

\section{Huygens Constructions}

The implications of Eqs.~(\ref{eq:vxtransinv}) and (\ref{eq:vytransinv}) may be found by comparing the results of two Huygens constructions, one for a ray normally incident on a block at rest, and the other for a ray normally incident on a moving block.

First we consider a Huygens construction in the rest frame $S'$ of the block.  A wavefront is produced by a ray entering the block at normal incidence, $\theta_1'=0$.  If the source of the expanding wavefront is placed at the origin of frame $S'$ at time $t'=0$, the equation of the wavefront at a later time $t'$ is
\begin{equation}
x'^2+y'^2=\left(\dfrac{c}{n}\right)^2t'^2.
\end{equation}
An identical wavefront may be drawn with its source displaced in the positive $x'$-direction.  A horizontal line drawn tangent to the wavefronts is a line of constant phase.  As time passes, this line moves vertically upward, in the $y'$-direction.  The wave vector $\vec{k}$ is perpendicular to this line.  The Poynting vector, $\vec{S}'$, drawn from the source to the point of tangency, is clearly parallel to the wave vector, $\vec{k}'$.  For the block at rest, $\vec{S}'$ and $\vec{k}'$ are both in the same direction, along the positive $y'$-axis.

\begin{figure}[h] 
  \centering
  \includegraphics[bb=0 0 481 199,width=5.67in,height=2.34in,keepaspectratio]{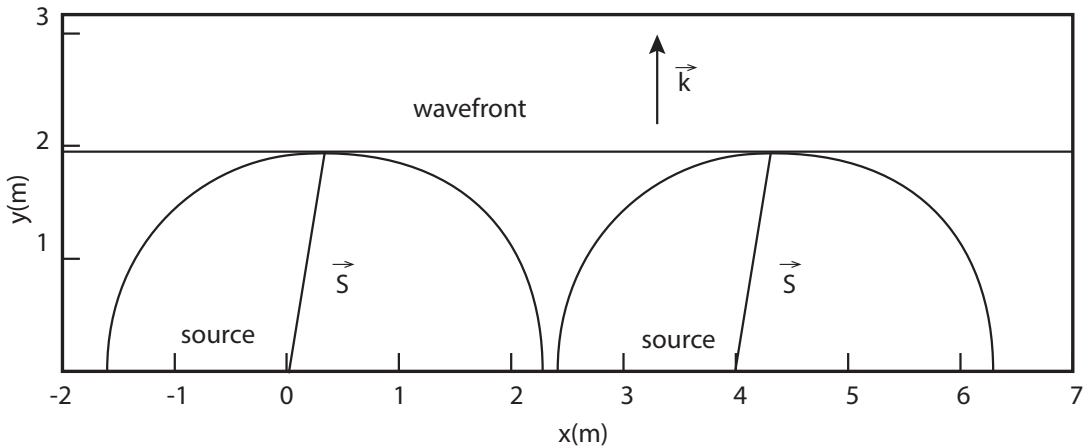}
  \caption{Two wavefronts spread out from two sources of light entering the bottom of the block at normal incidence, as seen in the lab frame $S$.  The horizontal line, the wavefront, is a line of constant phase that moves vertically through the block, in the direction of the wave vector $\vec{k}$.  The Poynting vector $\vec{S}$ is directed from each source to the point of tangency.  The situation shown is for 
$n=1.5$, $u/c=0.2$, and $t=10^{-8}$~s.}
  \label{fig:wavefronts}
\end{figure}

Now consider a Huygens construction in the lab frame $S$, in which the block is moving.  A wavefront is produced by a ray entering the block at normal incidence, $\theta_1=0$.  Assuming, as usual, that the origins of the frames $S$ and $S'$ coincide at $t = t' = 0$, the standard Lorentz transformations show that the coordinates of this wavefront in the moving block in lab frame $S$ must satisfy
\begin{equation}
\left(x-ut\right)^2+y^2\left(1-\dfrac{u^2}{c^2}\right)=\dfrac{1}{n^2}\left(ct-\dfrac{ux}{c}\right)^2.
\end{equation}
For a given value of $t$, the coordinates $(x,y)$ are the same as those obtained by using $x=v_{2x}t$ and $y=v_{2y}t$, where $v_{2x}$ and $v_{2y}$ are given by Eqs.~(\ref{eq:vxtransinv}) and (\ref{eq:vytransinv}), respectively.  An identical wavefront may be drawn with its source displaced in the positive $x$-direction.  Figure~\ref{fig:wavefronts} shows two of these wavefronts.  The horizontal line is tangent to the wavefronts, and so is a line of constant phase.  As time passes, this line moves vertically upward, in the $y$-direction.  The wave vector $\vec{k}$ is perpendicular to this line, in the $y$-direction, and so does not show a transverse relativistic drift.  However the Poynting vector $\vec{S}$, drawn from the source to the point of tangency, deviates in the positive $x$-direction.  The angle of deviation from the $y$-direction of the line connecting the source and tangent point is given by Eq.~(\ref{eq:drift}), and so shows a transverse relativistic drift.  This deviation of $\vec{k}$ and $\vec{S}$ is routinely found in anisotropic crystals; see, for example, Refs.\ \onlinecite{Leach} and \onlinecite{Hecht}.  The conclusion is that the wave vector $\vec{k}$ in the moving block shows how the wavefronts move and does not show a transverse relativistic drift, while the Poynting vector $\vec{S}$ in the moving block shows how the momentum moves and does show a transverse relativistic drift.

\section{The Lorentz Transformation of the Momentum-Energy Four-Vector}

We now consider the momentum of a photon in a transparent nondispersive dielectric medium.  When a photon arrives at the block's surface, it is either completely reflected or completely transmitted into the block.  We need consider only the transmitted photons.  
In the vacuum in either frame, $E=pc$, so 
\begin{equation}
E_1=p_1c\mbox{\ \ \ \ \ and\ \ \ \ \ }E_1'=p_1'c.
\label{eq:vacuumE}
\end{equation}

The momentum and energy of the refracted effective photon are given by Eqs.~(\ref{eq:effp}) and (\ref{eq:effE}), which in the present notation are
\begin{equation}
p_2'=n^\alpha p_1'
\label{eq:effpnew}
\end{equation}
and
\begin{equation}
E_2'=n^{\alpha +1}p_1'c=n^{\alpha +1}E_1'.
\label{eq:effEnew}
\end{equation}


The Lorentz transformations for the momentum-energy four-vector are
\begin{eqnarray}
p_x'&=&\gamma\left(p_x-\dfrac{uE}{c^2}\right)\\[1ex]
E'&=&\gamma\left(E-up_x\right).
\end{eqnarray}
 
We begin with a photon with momentum $p_1$ and energy $E_1$ as observed in lab frame $S$.  The ray is incident upon the bottom surface of the block, making an angle $\theta_1$ with the normal to the block's surface.  The $x$-component of the incident momentum is 
\begin{equation}
p_{1x}=p_1\sin\theta_1.\label{eq:p1x}
\end{equation}    
We now use the Lorentz transformations to find the components of the incident photon momentum in frame $S'$.  Employing Eq.~(\ref{eq:vacuumE}), we obtain
\begin{eqnarray} 
p_{1x}'&=&\gamma\left(p_{1x}-\dfrac{uE_1}{c^2}\right)=\gamma\left(p_1\sin\theta_1-\dfrac{up_1c}{c^2}\right)=\gamma p_1\left(\sin\theta_1-\dfrac{u}{c}\right)\\[1ex]
E_1'&=&\gamma\left(E_1-up_{1x}\right)=\gamma\left(p_1c-up_1\sin\theta_1\right)=\gamma p_1c\left(1-\dfrac{u}{c}\sin\theta_1\right).
\end{eqnarray} 

Next, using Snell's law, Eq.~(\ref{eq:Snell}) and Eq.~(\ref{eq:effpnew}) for the effective momentum in the block frame $S'$, we have
\begin{equation} 
p_2'\sin\theta_2'=\left(n^\alpha p_1'\right)\left(\dfrac{1}{n}\sin\theta_1'\right)=n^{\alpha -1}p_1'\sin\theta_1',
\end{equation}
or
\begin{equation}
p_{2x}'=n^{\alpha -1}p_{1x}'=n^{\alpha -1}\gamma p_1\left(\sin\theta_1-\dfrac{u}{c}\right).
\end{equation}
From Eq.~(\ref{eq:effEnew}) we have
\begin{equation}
E_2'=n^{\alpha +1}E_1'=n^{\alpha +1}\gamma p_1c\left(1-\dfrac{u}{c}\sin\theta_1\right).
\end{equation}
Thus
\begin{equation}
p_1'=\dfrac{E_1'}{c}=\gamma p_1\left(1-\dfrac{u}{c}\sin\theta_1\right).
\end{equation}
The $y'$-component of the photon's momentum inside the block in the rest frame $S'$ is then
\begin{equation}
p_{2y}'=\sqrt{p_2'^2-p_{2x}'^2}=\sqrt{\left(n^\alpha p_1'\right)^2-\left[n^{\alpha -1}\gamma p_1\left(\sin\theta_1-\dfrac{u}{c}\right)\right]^2}.
\end{equation}
Thus
\begin{equation}
p_{2y}'=\sqrt{n^{2\alpha}\gamma^2p_1^2\left(1-\dfrac{u}{c}\sin\theta_1\right)^2-n^{2\alpha -2}\gamma^2p_1^2\left(\sin\theta_1-\dfrac{u}{c}\right)^2}
\end{equation}
so that
\begin{equation}
p_{2y}'=n^\alpha\gamma p_1\sqrt{\left(1-\dfrac{u}{c}\sin\theta_1\right)^2-\dfrac{1}{n^2}\left(\sin\theta_1-\dfrac{u}{c}\right)^2}.
\end{equation}

Finally, we use the inverse Lorentz transformations,
\begin{eqnarray}
p_{2x}&=&\gamma\left(p_{2x}'+\dfrac{uE_2'}{c^2}\right)\\[1ex]
p_{2y}&=&p_{2y}'
\end{eqnarray} 
to obtain the components of the momentum-energy four-vector inside the moving block as measured from the lab frame $S$.  As before, we are really interested in $\tan\theta_2$,
\begin{equation}
\tan\theta_2=\dfrac{p_{2x}}{p_{2y}}=\gamma\,\dfrac{p_{2x}'+\dfrac{uE_2'}{c^2}}{p_{2y}'}.
\end{equation}
Therefore
\begin{equation}
\tan\theta_2=\gamma\,\dfrac{n^{\alpha -1}\left(\sin\theta_1-\dfrac{u}{c}\right)+n^{\alpha +1}\dfrac{u}{c}\left(1-\dfrac{u}{c}\sin\theta_1\right)}{n^\alpha\sqrt{\left(1-\dfrac{u}{c}\sin\theta_1\right)^2-\dfrac{1}{n^2}\left(\sin\theta_1-\dfrac{u}{c}\right)^2}}.
\end{equation}
A few lines of algebra produces the final result,
\begin{equation}
\tan\theta_2=\gamma\,\dfrac{\left(n^2-1\right)\dfrac{u}{c}+\left(1-\dfrac{n^2u^2}{c^2}\right)\sin\theta_1}{\sqrt{n^2\left(1-\dfrac{u}{c}\sin\theta_1\right)^2-\left(\sin\theta_1-\dfrac{u}{c}\right)^2}}\label{eq:tanth2p}.
\end{equation}
This is identical to Eq.~(\ref{eq:tanth2v}).  It shows that regardless of the value of $\alpha$, the effective photon follows the trajectory of the ray and yields the transverse relativistic drift that is experimentally observed.  This might have been expected; it is ultimately a result of the relativistic self-consistency of our definition of the effective photon.  We therefore must turn to another argument to determine the value of $\alpha$.

\section{Photon Energy Conservation and the Abraham Momentum}

Let's return to considering the dielectric block at rest.  When a ray of identical monochromatic photons is incident upon the surface of the block, the photons are either transmitted or reflected.  For our non-dispersive medium, the energy of the photons is conserved; the incident energy is equal to the sum of the transmitted and reflected energies.  The energy of the reflected photons is certainly conserved because neither the number of photons nor their frequency changes upon reflection.  Because none of the photons from the incident ray are absorbed, the energy of the transmitted photons must also be conserved.  Therefore, since the energy of the photon is conserved when it enters the block at rest, Eq.~(\ref{eq:effE}) requires that $\alpha = -1$.  We must conclude that, to conserve energy, the effective photon must have the Abraham momentum, $p_{\rm eff}=p/n$.

The reader may be concerned that we may have, in our definition of an effective photon, predetermined that its momentum must be the Abraham momentum.  Might an alternative definition result in the effective photon having the Minkowski momentum, $p_{\rm eff}=np$ ($\alpha=1$)?  

We will now show that such an alternative definition of an effective photon is unphysical.  Specifically, we will assume that the Minkowski effective photon has the Minkowski momentum, so 
\begin{equation}
p_2'=np_1',
\end{equation}
and that the energy of the photon does not change as it enters the block at rest,
\begin{equation}
E_2'=E_1'=p_1'c.
\end{equation}
Note that we do {\em not} require that the Minkowski effective photon move with speed $c/n$.  In this case, we can solve for the speed of the Minkowski effective photon in the block at rest using Eqs.~(\ref{eq:relE}) and (\ref{eq:relp}),
\begin{equation}
\dfrac{p_2'c}{E_2'}=\dfrac{v}{c}.
\end{equation}
Substituting for $p_2'$ and $E_2'$ shows that
\begin{equation}
\dfrac{v}{c}=\dfrac{np_1'c}{p_1'c}=n.
\end{equation}
The Minkowski effective photon must be superluminal, which is unphysical.  The mass of the Minkowski effective photon comes from Eq.~(\ref{eq:Epm}),
\begin{equation}
E_2'^2-p_2'^2c^2=m^2c^4.
\end{equation}
This quickly yields
\begin{equation}
m=\dfrac{p_1'}{c}\sqrt{1-n^2},
\end{equation}
which is imaginary, and so is also unphysical.  Nevertheless, if we insist on using this new formulation for the Minkowski effective photon, another problem arises.  Upon examining the form of the momentum four-vector, $(p_x, p_y, p_z, E/c)$, and comparing it to the wave four-vector, $(k_x, k_y, k_x, \omega/c)$, we realize that the momentum and wave four-vectors transform in the same way under a Lorentz transformation.  In the rest frame $S'$ we have $p_2'=np_1'$ and $k_2'=nk_1'$;  $E_1'=E_2'$ and $\omega_1'=\omega_2'$; and in the vacuum in both frames $E_1=p_1c$ and $\omega_1=k_1c$.  Indeed, we will arrive at Eq.~(\ref{eq:tanth2k}) for $\tan\theta_2$, which does not show a relativistic drift for normal incidence ($\theta_1=0$).  
Therefore we must conclude that the proposed definition of a Minkowski effective photon is not consistent with the experimentally observed relativistic drift.  This eliminates the alternative definition of the Minkowski effective photon.  Such an effective photon would be superluminal, have an imaginary effective mass, and not reproduce the observed transverse relativistic drift.  This reinforces our conclusion that the effective photon must have the Abraham momentum, $p_{\rm eff}=p/n$.

Finally, we note in passing that our result disagrees with that of Wang\cite{Wang}, who uses a relativistic argument to conclude that the Abraham momentum and energy ($\alpha = -1$) do not constitute a Lorentz four-vector in a dielectric medium.  However, Wang's analysis incorrectly assumes that the momentum $\vec{p}$ and the wave-vector $\vec{k}$ are in the same direction in a moving block.  We have shown in Section~II that Wang's assumption is invalid.

\section{Conclusions}
We have used relativistic arguments to define an effective photon in a transparent nondispersive dielectric medium at rest that follows the path of the refracted ray, transports the ray's momentum and energy, and moves with speed $c/n$.  The effective photon has a momentum given by Eq.~(\ref{eq:effp}) and an energy given by Eq.~(\ref{eq:effE}).  Using a Lorentz transformation of the wave four-vector, we have demonstrated that the Poynting vector $\vec{S}$ and wave vector $\vec{k}$ diverge and are not in the same direction in a moving dielectric medium.  

For an effective photon with its momentum in a dielectric medium given by Eq.~(\ref{eq:effp}), we calculated the transverse relativistic drift and showed that the transverse drift is independent of the value of $\alpha$.  However, imposing the constraint of conservation of energy, we have shown that, in a dielectric medium, $\alpha=-1$ and that the effective photon carries momentum $p_{\rm eff}=p/n$ (the Abraham momentum).  Furthermore, we have shown that adopting the Minkowski momentum for an alternative definition of an effective photon leads to results which are unphysical and do not display the experimentally observed transverse relativistic drift.

\begin{acknowledgments}


\end{acknowledgments}

\end{document}